\documentclass[doublecol]{epl2} % for 2 columns style with line numbers
\usepackage{hyperref}

\newcommand{\be}{\begin{equation}}
\newcommand{\ee}{\end{equation}}

\title{Dynamical quantum phase transitions: a brief survey}

\author{Markus Heyl\inst{1}}
\shortauthor{M. Heyl}
\institute{                    
  \inst{1} Max-Planck-Institut f\"ur Physik komplexer Systeme, N\"othnitzer Str. 38, 01187 Dresden, Germany
}


\abstract{
Nonequilibrium states of closed quantum many-body systems defy a thermodynamic description.
As a consequence, constraints such as the principle of equal a priori probabilities in the microcanonical ensemble can be relaxed, which can lead to quantum states with novel properties of genuine nonequilibrium nature.
In turn, for the theoretical description it is in general not sufficient to understand nonequilibrium dynamics on the basis of the properties of the involved Hamiltonians.
Instead it becomes important to characterize time-evolution operators which adds time as an additional scale to the problem.
In these Perspectives we summarize recent progress in the field of dynamical quantum phase transitions, which are phase transitions in time with temporal nonanalyticities in matrix elements of the time-evolution operator.
These transitions are not driven by an external control parameter, but rather occur due to sharp internal changes generated solely by unitary real-time dynamics.
We discuss the obtained insights on general properties of dynamical quantum phase transitions, their physical interpretation, potential future research directions, as well as recent experimental observations.
}

\begin{document}

%% here a revision

%\revision{Insert here the text.
%See fig.~\ref{fig.1}, table~\ref{tab.1} and eq.~(\ref{eq.1}).
%See also~\cite{b.a,b.b}.}

\maketitle

\section{Introduction}

\begin{figure}
	\centering
	\includegraphics[width=0.9\columnwidth]{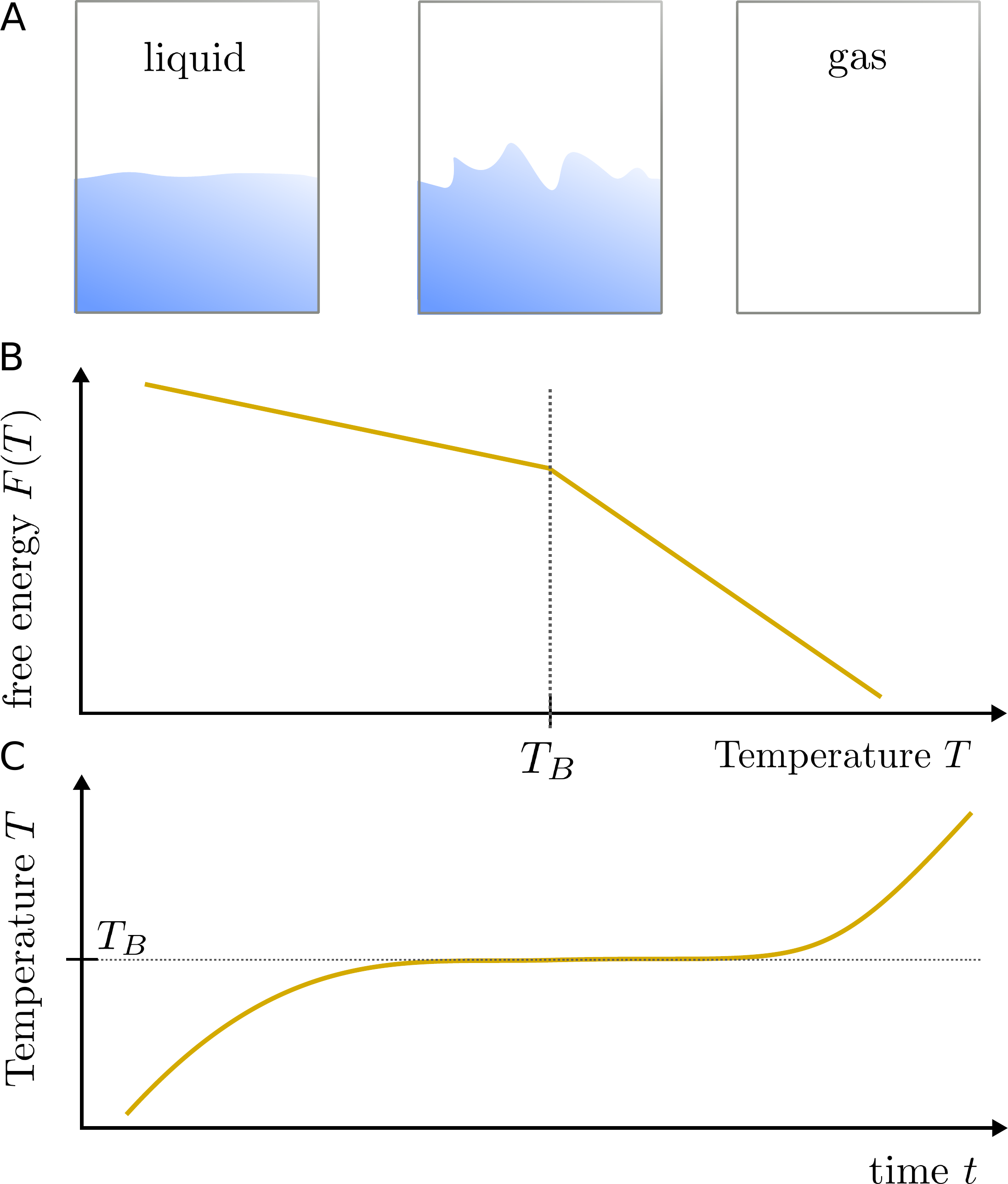}
	\caption{{\bf A} Schematic illustration of the first-order boiling transition of water. {\bf B} The free energy $F(T)$ exhibits nonanalytic behavior as a function of the external control parameter $T$ at the boiling temperature $T_B$. {\bf C} When monitored as a function of real time the temperature $T$ changes smoothly upon heating up the liquid.}
	\label{fig.1}
\end{figure}

We encounter phase transitions under various incarnations in every-day life, such as in the case of boiling water or the melting of ice.
From a thermodynamic perspective, these processes manifest as a nonanalytic behavior upon tuning an externally-controlled parameter~\cite{1987imsm.book.....C,1998AmJPh..66..164C,2011qpt..book.....S}, such as the temperature.
In the case of boiling water for instance the free energy $F(T)$ exhibits a nonanalytic structure upon increasing the temperature $T$, as illustrated in Fig.~\ref{fig.1}B.
However, all these processes take their time.
Suppose that we monitor the temperature $T$ during the actual heating process in real time, then we would observe the behavior displayed in Fig.~\ref{fig.1}C.
After an initial growth, $T$ levels off to the boiling temperature $T_B$.
A further increase can only occur, when the heating mechanism has provided the latent heat necessary to cross this first-order phase transition.
And since a realistic heating machine can only provide a finite power, this takes time.
Importantly, monitoring the temporal heating process of crossing the transition is now smooth in contrast to the characteristic nonanalytic behavior as a function of the control parameter.

\begin{figure}
	\centering
	\includegraphics[width=\columnwidth]{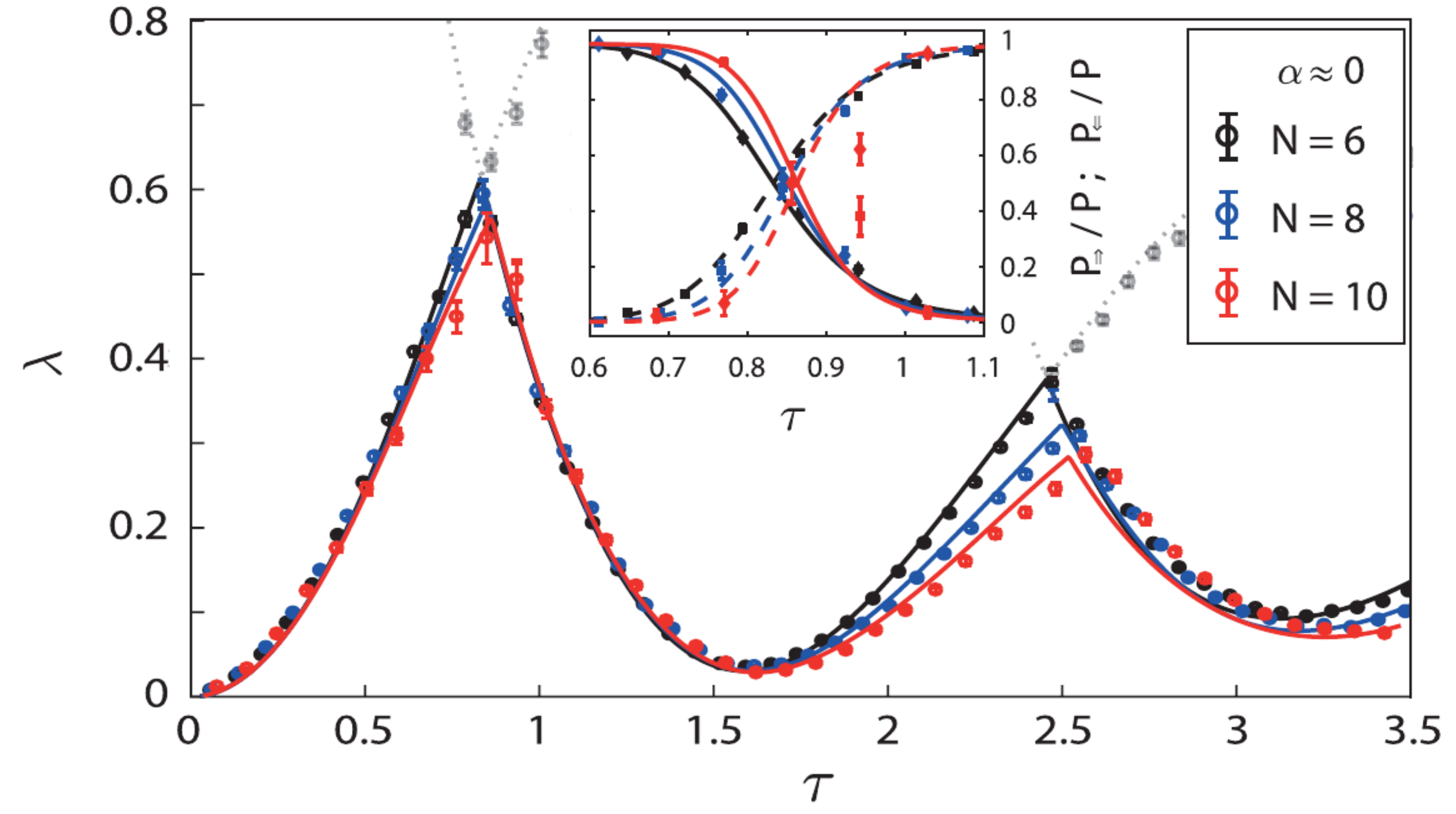}
	\caption{At a dynamical quantum phase transition physical quantities can become nonanalytic as a function of time. Here, the data from a trapped ion experiment shows that the Loschmidt echo rate function $\lambda(t)$ for a long-range transverse-field Ising model can exhibit kinks as a function of a rescaled dimensionless time $\tau$~\cite{2017PhRvL.119h0501J}.}
	\label{fig.2}
\end{figure}

In the quantum world, remarkably, this can be different: a quantum many-body system can undergo a dynamical quantum phase transition (DQPT) with physical quantities becoming nonanalytic as a function of real time~\cite{2013PhRvL.110m5704H,2018RPPh...81e4001H}, see Fig.~\ref{fig.2} for some recently obtained experimental measurements.
At such a DQPT a system can therefore show nonsmooth properties caused solely by drastic internal changes and not imposed by the exterior.
In this article we give an overview over this phenomenon including both a summary of the theoretical and experimental developments as well as a discussion on open challenges and future prospects.

These DQPTs occur in closed quantum many-body systems during unitary real-time evolution, so that the influence of an environment can be neglected.
In addition to fundamental theoretical aspects~\cite{2011RvMP...83..863P, 2015NatPh..11..124E, 2015ARCMP...6..383A, 2015ARCMP...6...15N, 2017NatPh..13..424M } the motivation to study such nonequilibrium unitary dynamics originates to a large extent in experimental advances in so-called quantum simulators over the last years~\cite{2008RvMP...80..885B, 2012NatPh...8..267B, 2012NatPh...8..277B, 2014RvMP...86..153G}.
In platforms such as ultra-cold atoms or trapped ions among many others it has become possible to experimentally realize and probe such scenarios with a high degree of control and precision, which has lead to the observation of inherent dynamical phenomena without equilibrium counterparts.
This includes the observation of many-body localization~\cite{2015Sci...349..842S, 2016Sci...352.1547C, 2016NatPh..12..907S}, prethermalization~\cite{2012Sci...337.1318G,2017SciA....3E0672N,2018arXiv180905554S}, particle production in lattice gauge theories~\cite{2016Natur.534..516M}, discrete time crystals~\cite{2017Natur.543..221C, 2017Natur.543..217Z}, or dynamical quantum phase transitions~\cite{2017PhRvL.119h0501J, 2018NatPh..14..265F, 2017Natur.551..601Z, 2018arXiv180609269G, 2018arXiv180704483T, 2018arXiv180610871W, 2018arXiv180803930X}.
The theoretical description and understanding of such nonequilibrium quantum states is, however, facing major challenges.
On a fundamental level, these states don't exhibit a description in terms of a free energy and are therefore beyond thermodynamics.
This, however, might not only be seen as an obstacle but rather as an opportunity to generate new quantum states beyond equilibrium constraints such as the principle of equal a priori probabilities in the microcanonical ensemble.
As a consequence the field of nonequilibrium quantum physics admits a framework to generate many-body states with novel properties impossible to realize within conventional thermodynamics.
A prominent example in this direction constitutes the celebrated time crystal~\cite{2016PhRvL.116y0401K, 2016PhRvL.117i0402E, 2017NatPh..13..424M}, which cannot exist in equilibrium states~\cite{2013PhRvL.110k8901B,2015PhRvL.114y1603W}.
From an alternative point of view the main challenge in the understanding of nonequilibrium quantum states is that now it is not sufficient to understand the properties of Hamiltonians.
Instead, we have to characterize time-evolution operators:
\be
	U(t) = \mathcal{T}e^{-i \int_0^t dt' \, H(t')}\, .
\ee
Here, $\mathcal{T}$ denotes the time-ordering prescription and $H(t)$ the, in general, time-dependent Hamiltonian.
Crucially, the propagator $U(t)$ contains one additional scale, which is time itself.
As we will discuss in the remainder of these Perspectives, this additional scale can lead to new physics.
However, it remains a central question how to extract general principles in time-evolution operators, i.e., quantum dynamics, and to which extent we can describe nonequilibrium quantum many-body states in a unified manner.
It is one of the main purpose of this article to summarize how the theory of DQPTs can contribute to this field.

\section{Dynamical quantum phase transitions}

A DQPT is a phase transition in time driven by sharp internal changes in the properties of a quantum many-body state and not driven by an external control parameter such as temperature or pressure.
The central object for the theory of DQPTs is the so-called Loschmidt amplitude
\be
	\mathcal{G}(t) = \langle \psi_0 | \psi_0(t)\rangle =  \langle \psi_0| U(t) | \psi_0 \rangle\, , 
	\label{eq:def_G}
\ee
denoting the overlap between the initial state $|\psi_0\rangle$ and the time-evolved one $|\psi_0(t)\rangle$.
Alternatively, one can interpret $\mathcal{G}(t)$ as a matrix element of the time-evolution operator $U(t)$.
In this way, the study of $\mathcal{G}(t)$ serves the purpose of characterizing $U(t)$ instead of Hamiltonians and their respective thermal states, as is done in equilibrium.
In the following we will consider for convenience a specific nonequilibrium protocol, a so-called quantum quench.
For a quantum quench the initial condition $|\psi_0\rangle$ is chosen as the ground state of an initial Hamiltonian $H_0$ and the Hamiltonian driving the time evolution as $H(t)=H$.
Such a scenario results from a sudden switching of a system parameter at time $t=0$. 
The Loschmidt amplitude $\mathcal{G}(t)$ then acquires the following form
\be
	\mathcal{G}(t) =  \langle \psi_0 | e^{-iHt} |\psi_0 \rangle \, .
\ee
While we will restrict the discussion to such quantum quenches, the theory of DQPTs is formulated in a much more general context~\cite{2018RPPh...81e4001H}, as one can already see from Eq.~(\ref{eq:def_G}), which does not rely on a specific nonequilibrium protocol.
Although we will focus here on the case of initial pure states, extensions to mixed states have been discussed in the literature recently~\cite{2016PhRvB..93j4302A,2017PhRvB..96r0303B,2017PhRvB..96r0304H,2018PhRvB..97q4401L,2018NatSR...811921B}.
While real-time nonanalyticities in the Loschmidt echo have been first found in Ref.~\cite{2010PhRvE..81b0101P}, the interpretation as a DQPT and the connection to conventional phase transitions has been formulated in Ref.~\cite{2013PhRvL.110m5704H}.

It is one of the most important insights that Loschmidt amplitudes on a formal level resemble partition functions in conventional statistical physics~\cite{2013PhRvL.110m5704H}.
While this might appear already evident from, e.g. the defining equation of the partition function $Z$ of the canonical ensemble $Z=\mathrm{Tr} e^{-\beta H}$, there is an even stronger formal analogy to so-called boundary partition functions $Z_B$ which have the structure $Z_B=\langle \psi_A | e^{-RH} | \psi_B \rangle$~\cite{1995NuPhB.453..581L}.
This class of partition functions describes systems subject to boundary conditions on two ends, which are encoded in the boundary states $|\psi_{A/B}\rangle$.
The spatial distance between the two boundaries is proportional to $R$ and $H$ denotes the bulk Hamiltonian.
In this context one can interpret the Loschmidt amplitude $\mathcal{G}(t)$ as a boundary partition function in time with $|\psi_0\rangle$ implementing the respective temporal boundary conditions.
As the free energy corresponding to an equilibrium partition function becomes nonanalytic at a phase transition, so can therefore
\be
	g(t) = -\frac{1}{N} \log \big[ \mathcal{G}(t) \big] \, ,
\ee
which can be viewed as a dynamical counterpart to a free energy density up to a differently chosen normalization with $N$ denoting the number of degrees of freedom.
A point in time $t^\ast$, where such a nonanalyticity occurs, we define in the following as a DQPT.
It will be useful to also consider the quantity
\be
	\lambda(t) = -\frac{1}{N} \log \big[ \mathcal{L}(t) \big] \, , \quad \mathcal{L}(t) = \big| \mathcal{G}(t) \big|^2 \, ,
\ee
which is the analog of $g(t)$ for the probability $\mathcal{L}(t)$ associated with $\mathcal{G}(t)$.
Evidently, $\lambda(t) = 2 \mathrm{Re}[g(t)]$ such that a nonanalytic behavior in $g(t)$ directly translates into a nonanalytic behavior in $\lambda(t)$.
In Fig.~\ref{fig.2} one can see a DQPT in $\lambda(t)$ for the paradigmatic Ising model upon quenching a transverse field.
In the meantime DQPTs have been observed in a variety of different platforms addressing diverse aspects and physical phenomena. This includes experiments in ultra-cold atoms in optical lattices~\cite{2018NatPh..14..265F}, trapped ions~\cite{2017PhRvL.119h0501J}, quantum walks~\cite{2018arXiv180610871W, 2018arXiv180803930X}, nanomechanical oscillators~\cite{2018arXiv180704483T}, and superconducting qubits~\cite{2018arXiv180609269G}.
It will be the goal of the remainder of this article to discuss the physical meaning of these real-time nonanalyticities and to explore their implications for the understanding of the dynamics in quantum many-body systems.

\section{General implications}

While the real-time nonanalyticities themselves might be regarded already as an intriguing phenomenon, it remains a central question to understand the physical implications of DQPTs.
On a more general note, let us emphasize that the dynamical analog $g(t)$ to the free energy density is not a thermodynamic potential.
In particular, derivatives of $g(t)$ are not related to measurable quantities as compared to the equilibrium case where, for example, the second derivative of the free energy with respect to temperature yields the specific heat.
As a consequence, the observation of a temporal nonanalyticity in $g(t)$ does not immediately imply a measurable signature in a physical observable.
A DQPT rather indicates a point in time where the time-evolution operator $U(t)$ exhibits a drastic change in its properties without providing insights into the character of this changes, which requires a further analysis.
Let us take the chance at this point to draw an analogy to an equilibrium scenario sharing some similarities.
Quantum phase transitions for 1D systems can be detected through those points in parameter space where the area law of the entanglement entropy is violated~\cite{2003PhRvL..90v7902V}.
This method for detection has found various applications in recent years since it represents a system-independent general-purpose tool that doesn't require detailed knowledge about the physical system.
The study of the entanglement entropy alone, however, does not provide the full physical picture, for which we still need to identify the respective order parameters for instance.
In a similar way, a DQPT implies a point in time with a radical change in the time-evolution operator and therefore represents an analogous system-independent indicator.
The nonanalytic temporal behavior in $g(t)$, however,  does not specify the physical origin of the DQPT, which requires further analysis.
These considerations naturally lead to the question: what do we learn from DQPTs then?
Clearly, this field is still developing and new facets are likely to explored in the future.
However, currently, two major aspects are worthwhile mentioning here:
(i) it is, remarkably, possible for some classes of models to obtain information about ground state phase diagrams from the study of DQPTs although being driven far away from equilibrium;
(ii) the theory of DQPT provides general principles of quantum dynamics, which allow us to understand classes of nonequilibrium scenarios instead of analyzing individual problems.

\section{Relation to underlying equilibrium phase transitions}
In many cases it has been recognized that DQPTs are directly connected to the underlying equilibrium phase transitions of the considered models.
However, it has turned out that this connection is, in general, not to be understood as a one-to-one correspondence~\cite{2014PhRvB..89p1105V, 2014PhRvB..89l5120A, 2015PhRvB..92g5114S, 2017PhRvB..96m4427H}.
Therefore, DQPTs constitute a genuine nonequilibrium phenomenon.
The relation between the appearance of DQPTs and the underlying ground state properties of the Hamiltonian are most extensively understood for topological two-band models~\cite{2015PhRvB..91o5127V,2016PhRvB..93h5416B,2016PhRvL.117h6802H}.
A quantum quench across a topological ground state phase transition in one dimension (1D) always leads to a DQPT in real-time evolution, which is why these DQPTs are also termed topologically protected~\cite{2015PhRvB..91o5127V, 2016PhRvL.117h6802H}.
The reverse, however, is not always true.
It can happen that a system undergoes a DQPT without crossing an underlying equilibrium transition.
These DQPTs are therefore 'accidental' and upon smoothly changing the Hamiltonian parameters, they can be made disappear without closing a gap along the way.
This strong connection between equilibrium critical points and DQPTs can be used in 1D, remarkably, to map out ground state phase diagrams based just on the study of DQPTs~\cite{2016PhRvB..93h5416B}.
In this way, information about ground states can be inferred via nonequilibrium dynamics, which necessarily takes place at elevated energy densities.
Specifically, the dynamics of a dynamical topological order parameter~\cite{2016PhRvB..93h5416B}, which can be defined on general grounds for these models, is capable to distinguish uniquely, whether an underlying quantum phase transitions has been crossed or not.
This has, for example, be used in split-step quantum walks to experimentally obtain information about phase boundaries~\cite{2018arXiv180803930X}.
For systems other than the topological ones discussed in the paragraph before, it is not possible to rigorously connect equilibrium and dynamical phase transitions on general grounds.
For many models, however, the appearance of DQPTs is nevertheless directly related, so that also here DQPTs can be utilized to map out equilibrium ground state phase diagrams, by keeping in mind, however, that this might be subject to change upon continuously deforming the Hamiltonian even without gap closing.
Importantly, a weak symmetry-preserving perturbation is, however, not sufficient to achieve this, see also the discussion on the robustness of DQPTs below.
Finally, there are also models for which quantum quenches across equilibrium transitions do not lead to DQPTs because of kinetic constraints~\cite{2014PhRvB..89p1105V, 2014PhRvB..89l5120A}.
The discussion up to now has only considered the case of ground state phase transitions.
How transitions at nonzero temperature reflect dynamically in DQPTs is currently much less known, in particular because such transitions require at least 2D, which is much more challenging to theoretically address with some notable exceptions~\cite{2014PhRvL.113z5702C, 2015PhRvL.115n0602H, 2018PhRvL.120m0601Z, 2017PhRvB..96m4427H, 2017PhRvB..96m4313W, 2017PhRvL.119h0501J,2018arXiv180807874H}.

\begin{figure}
	\centering
	\includegraphics[width=\columnwidth]{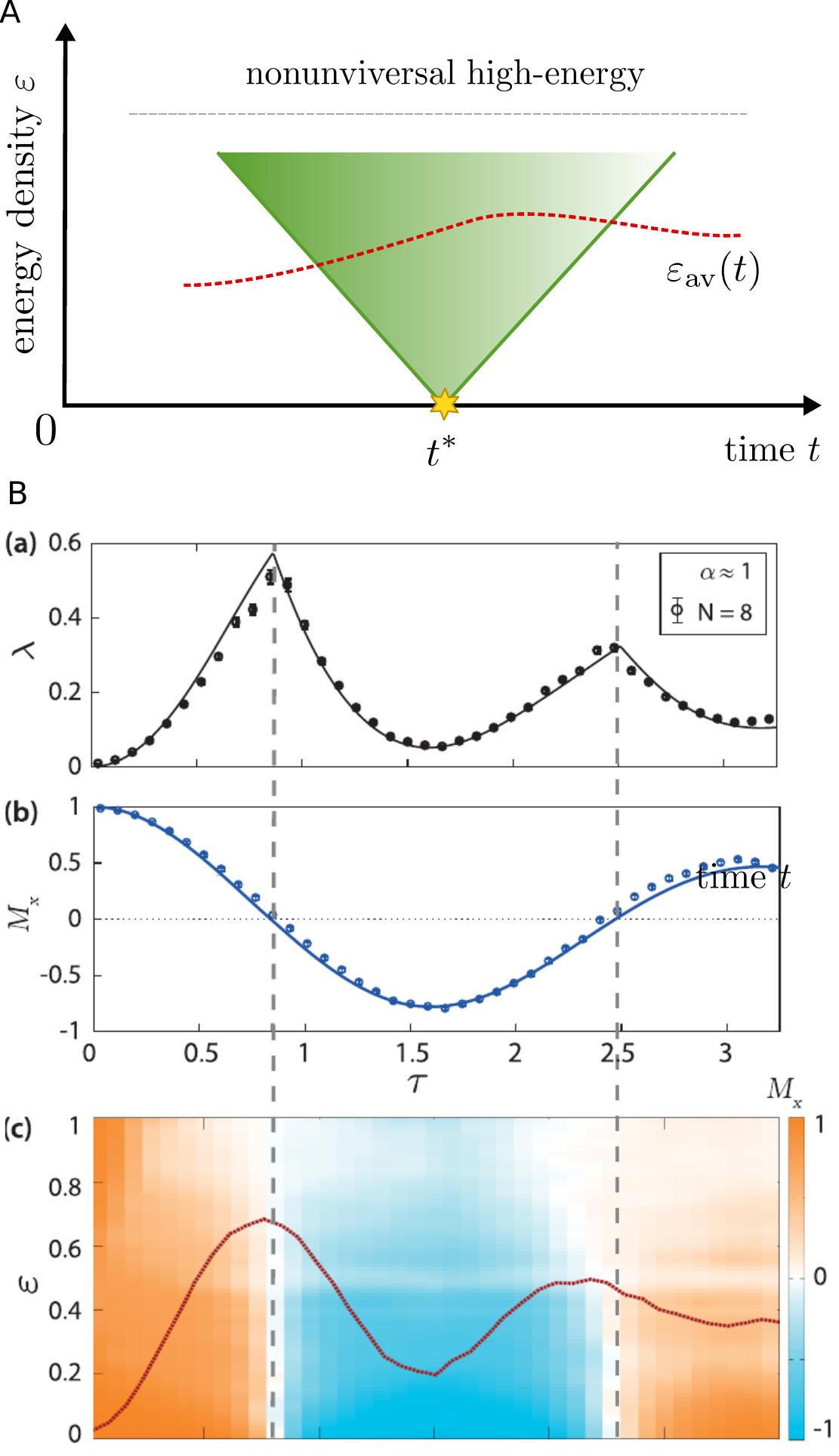}
	\caption{Dynamical counterpart to quantum critical regions in the vicinity of DQPTs. {\bf A} Schematic illustration of the energy density-time plane, where energy is measured with respect to the initial Hamiltonian and $\varepsilon=0$ corresponds to the ground state energy density. A DQPT occurs along $\varepsilon=0$ at a critical time $t^\ast$. The nonanalytic properties of the DQPT at $t^\ast$ can extend to $\varepsilon>0$, schematically depicted here as the green area. The average energy $\varepsilon_\mathrm{av}(t)$ is included as the dotted red line.
		{\bf B} Such dynamical analogs of quantum critical regions can, for some specific cases, be measured experimentally~\cite{2017PhRvL.119h0501J}, see {\bf (c)} where  the energy- and time-resolved magnetization $M_x(\varepsilon,t)$ is shown on a color scale.
		The dashed gray lines indicate the locations of two subsequent DQPTs, occurring in the unitary dynamics of realized long-ranged transverse-field Ising chain, see {\bf (a)}.
		In the vicinity of both of the two DQPTs an area, which is controlled by the DQPT (white), extends to $\varepsilon>0$ and intersects $\varepsilon_\mathrm{av}(t)$, which allows to conclude that these DQPTs control the nonequilibrium dynamics of the considered magnetization, as one can indeed observe as vanishing values for the mean magnetization $M_x$ in {\bf (b)}.
	}
	\label{fig.4}
\end{figure}

\section{Dynamical analogs to quantum critical regions}
The identification of DQPTs as nonequilibrium phase transitions has been argued initially on the basis of the formal similarity of the Loschmidt amplitude with equilibrium partition functions.
However, it is important to note again that $g(t)$ does not represent a thermodynamic potential despite of the formal similarities to free energy densities.
Derivatives of $g(t)$ cannot be connected directly to observables or correlation functions, leading immediately to the following open questions:
(i) Is there an indirect connection to observables or correlation functions then?
(ii) What do these DQPTs mean when the quantity $g(t)$, which entails the defining real-time nonanalyticities, does not function as a thermodynamic potential?
While many models and nonequilibrium protocols have been identified, where temporal nonanalyticities occur, for the understanding of DQPTs these two questions still remain as major challenges.

In the following, a physical interpretation of DQPTs, useful in many cases, will be summarized.
This interpretation is based on viewing the Loschmidt amplitude and echo as measures to probe the time-evolved quantum many-body state $|\psi_0(t)\rangle$ in the ground state manifold of the initial Hamiltonian $H_0$, since both $\mathcal{G}(t)$ and $\mathcal{L}(t)$ project $|\psi_0(t)\rangle$ back onto the ground state $|\psi_0\rangle$ of $H_0$.
Along these lines, let us decompose $|\psi_0(t)\rangle$ in the full eigenbasis $|\psi_\nu\rangle$ of the initial Hamiltonian:
\be
|\psi_0(t)\rangle = \sum_{\nu} a_\nu(t) | \psi_\nu \rangle \, , \quad a_\nu(t) = \langle \psi_\nu | \psi_0(t) \rangle \, ,
\ee
where $\nu=0$ corresponds to the initial ground state, i.e., $a_{\nu=0}(t) = \mathcal{G}(t)$.
From the perspective of this expansion $\mathcal{G}(t)$ quantifies one of the exponentially many amplitudes $a_\nu(t)$.
Thus, how can this single overlap be important, when most of the weights $a_\nu(t)$ are in $\nu\not=0$?
This can only be the case when $\nu=0$ does not represent a singular point, but rather when the properties of $\mathcal{G}(t)=a_0(t)$ extend to other amplitudes $\nu\not=0$.
An important example in equilibrium, where the properties of a single quantum many-body state extend to a significant portion of Hilbert space, is that of a quantum phase transition~\cite{2011qpt..book.....S}.
Although nonanalyticities as a function of a control parameter can be found only in the ground state, the zero-temperature critical point controls the whole quantum critical region in the temperature-control parameter plane.
It is the goal of the following discussion to argue that an analog to a quantum critical region can also exist for DQPTs, see also Fig.~\ref{fig.4}.
From an operational point of view, $\mathcal{L}(t) = | a_0 (t)|^2$ is the result of a projective measurement of the energy $E$ with the initial Hamiltonian $H_0$, where as a measurement outcome we have obtained $E=0$ upon choosing the zero of energy accordingly.
We might, however, also consider other possible measurement outcomes and probe the state's properties also at excited energy densities $\varepsilon=E/N>0$.
In case $a_0(t)$ does not represent a singular point, the DQPT occurring at $t=t^\ast$ along $\varepsilon=0$ extends also to $\varepsilon>0$, as illustrated in Fig.~\ref{fig.4}A.
Accordingly, the temperature-control parameter plane is replaced by an energy density-time plane with a potential dynamical analog to a quantum critical region.
Importantly, this can be measured even experimentally~\cite{2017PhRvL.119h0501J}, see also Fig.~\ref{fig.4}.

Despite of the various suggestive analogies to equilibrium quantum critical regions, there remains one central difference.
In equilibrium the temperature can, at least in principle, be chosen at will.
In the dynamical case discussed here this is not the case for the energy density.
Expectation values of local observables rather get their dominant contributions only from a limited set of states with energy densities close to the mean value $\varepsilon_\mathrm{av}(t) = \int d\varepsilon \, \varepsilon \, P(\varepsilon;t)$ due to a central limit theorem~\cite{2014PhRvL.113t5701H}, where $P(\varepsilon;t)$ denotes the energy density distribution function at time $t$.
In this way the energy density, at which we probe our system with local observables, is fixed by the nonequilibrium protocol itself via $\varepsilon_\mathrm{av}(t)$.
Whenever $\varepsilon_\mathrm{av}(t)$ crosses the dynamical analog of the critical region, the green area in Fig.~\ref{fig.4}A, one can expect that the underlying DQPT controls also the dynamics of local observables.
It might, however, also occur that $\varepsilon_\mathrm{av}(t)$ enters a non-universal regime at elevated energy densities, where an underlying DQPT might then not have a significant influence on observables.
Importantly, such dynamical analogs of quantum critical regions can be computed~\cite{2014PhRvL.113t5701H} and even measured~\cite{2017PhRvL.119h0501J}, see also Fig.~\ref{fig.4}B.
However, these quantitative calculations can only be done for fine-tuned models~\cite{2014PhRvL.113t5701H}.
While the picture from Fig.~\ref{fig.4}A is not tied to these particular problems, it remains a significant challenge for the future to develop a more general framework to quantitatively compute such critical regions.
Such a framework might be also particularly important for exploring whether other major properties of quantum critical regions take over to the dynamical case, such as scaling, for example.
For the models, in which critical regions as in Fig.~\ref{fig.4} can be computed, the question of scaling cannot be unambiguously addressed, since the exponents associated with the DQPT are all integer-valued and therefore can only hardly be distinguished from a trivial scaling.

\section{General properties}
The real-time nonanalyticities of DQPTs and the formal similarity of Loschmidt amplitudes to complex partition functions have initially motivated the notion of a dynamical quantum phase transition.
However, it is important to emphasize that equilibrium phase transitions are much more than just nonanalytic behavior.
For example, a continuous equilibrium transition separates two phases characterized by an order parameter, and inherits the powerful concepts of scaling and universality, which allow for a macroscopic description independent of microscopic details.
In the following we summarize the progress of the theory of DQPTs in connecting to such important equilibrium concepts.

\subsection{Dynamical order parameters}
\begin{figure}
	\includegraphics[width=\columnwidth]{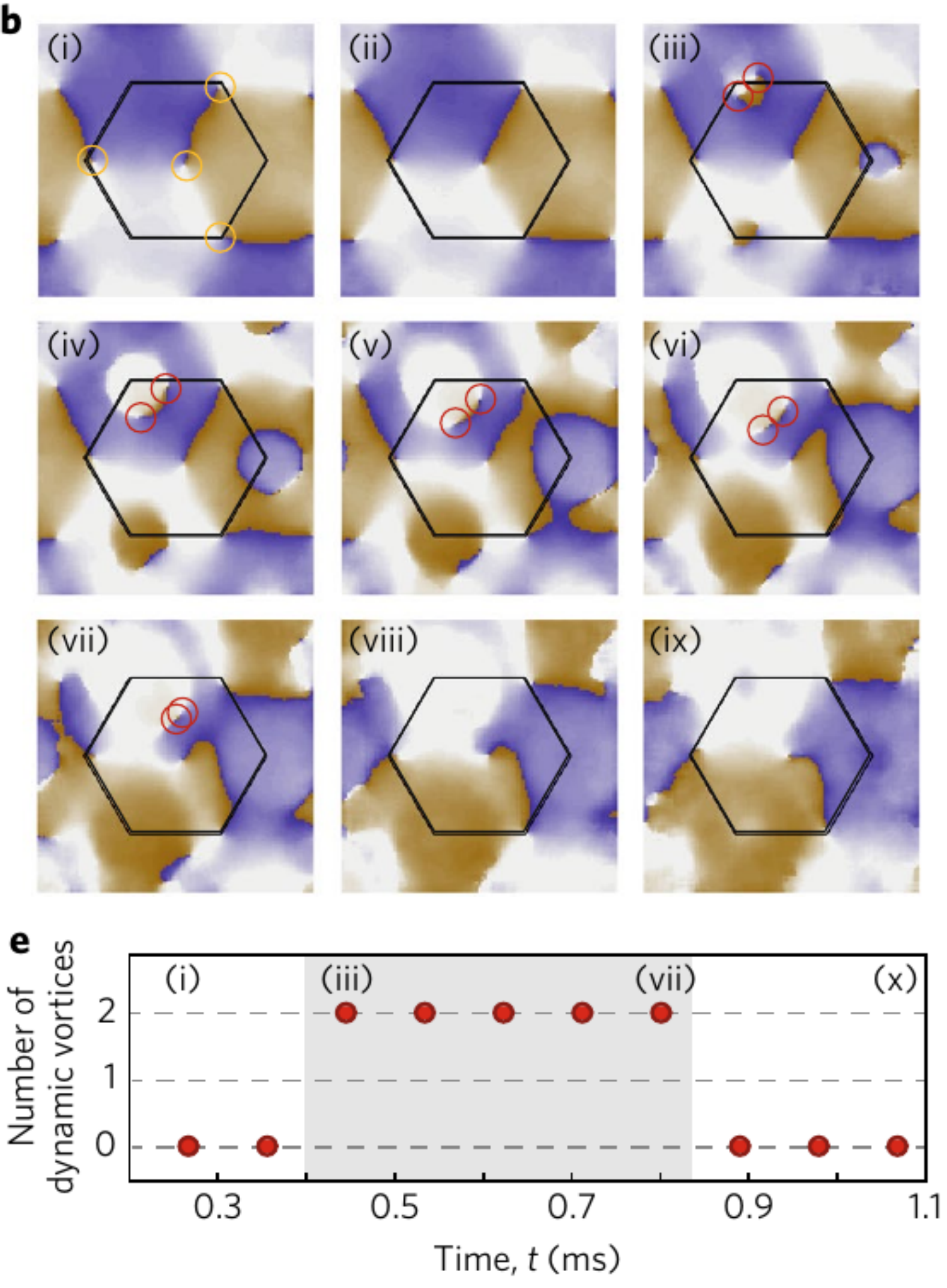}
	\caption{Observation of a dynamical order parameter in an ultra-cold atom experiment~\cite{2018NatPh..14..265F}. {\bf(b)} Subsequent snapshots of a momentum-dependent phase profile across the 2D Brioullin zone for increasing times from (i) to (ix). At some points in time, here between (ii) and (iii), suddenly pairs of vortices appear enclosed by the red circles. At a later time between (vii) and (viii), they recombine and annihilate. {\bf (e)} The number of dynamically generated vortices constitutes a dynamical order parameter for DQPTs happening at those points in time where the vortices a created and annihilated.}
	\label{fig.3}
\end{figure}
Order parameters for DQPTs have been reported for noninteracting topological two-band models both in 1D and 2D~\cite{2016PhRvB..93h5416B, 2017PhRvB..96r0303B, 2017PhRvB..96a4302B, 2017PhRvB..96r0304H, 2018NatPh..14..265F}.
These dynamical order parameters are topological quantum numbers detecting topological defects in phase profiles across the Brioullin zone.
Importantly, these have been measured in various different platforms~\cite{2018NatPh..14..265F, 2018arXiv180610871W, 2018arXiv180704483T, 2018arXiv180803930X}, one example is shown in Fig.~\ref{fig.3}.
Recently, a dynamical order parameter in a momentum-time plane of Green's functions has been reported in the context of gauge theories, which can be used for interacting and noninteracting systems on equal footing~\cite{2018arXiv180807885Z}.
All of these mentioned examples constitute order parameters of nonlocal nature associating a global topological quantum number to the time-evolved nonequilibrium wave function.
To which extent also local order parameters can exist is currently unknown.
However, it is clear from general principles that certain scenarios are impossible due to constraints induced by locality and causality.
Since DQPTs occur at a finite time, long-ranged quantum correlations cannot develop at a DQPT due to Lieb-Robinson bounds~\cite{1972CMaPh..28..251L, 2006JSP...124....1N, 2006CMaPh.265..781H}.
And therefore a conventional local order parameter associated to such long-range correlations cannot be formulated.
However, this leaves open the possibility still for order parameters which are non-local or global, as discussed for the topological systems and gauge theories before.
In this context let us note that non-local order parameters can also be identified for some quantum quenches in systems with symmetry-broken phases in equilibrium~\cite{2014PhRvL.113t5701H, 2018PhRvL.120m0601Z, 2017PhRvB..96m4313W, 2018arXiv180807874H} after a non-local projection onto a suitable low-energy manifold~\cite{2014PhRvL.113t5701H}.
Let us also take the chance at this point to repeat the saying that "choosing an order parameter is an art"~\cite{1992cond.mat..4009S}.
Having found a suitable one, implies a high degree of understanding of the studied transition, which, on general level, has not yet been achieved for DQPTs as compared to the equilibrium case.
This leaves open a promising avenue for future developments in this field.

\subsection{Scaling and universality}
Continuous phase transitions in equilibrium are associated with a divergent correlation length, as a consequence of which macroscopic properties become independent of microscopic details~\cite{1987imsm.book.....C,1998AmJPh..66..164C,2011qpt..book.....S}.
Whether such universal behavior can be found also for DQPTs is not known on a general level.
However, for a transverse-field Ising chain it has been established rigorously that the DQPTs appearing in this model are associated with an unstable fixed point of an exact renormalization group transformation (RG)~\cite{2015PhRvL.115n0602H}.
Thus, in the very equilibrium sense, this DQPT results from a divergent correlation length appearing in the Loschmidt amplitude.
Note, that this should not be confused with a divergent length scale in correlation functions, although these also show indicative features of the DQPT in their dynamics for these models~\cite{2015PhRvL.115n0602H,2018ScPP....4...13S}.
Further, for DQPTs in a 2D transverse-field Ising model strong indications of a divergent correlation length have been found.
Specifically, the nonanalytic real-time behavior follows the same scaling as for the equilibrium nonzero-temperature critical point of the 2D classical Ising model~\cite{2015PhRvL.115n0602H}.
While this suggests that scaling and universality also hold for this case, it has up to now not been possible to settle this rigorously, for example, by identifying the corresponding RG fixed point.

\subsection{Landau and effective field theories}
The universal properties at continuous equilibrium phase transitions are determined solely by macroscopic properties such as symmetries or dimensionality.
These are sufficient to construct Landau or effective field theories, which describe the universal properties in the vicinity of the transition.
Although for one particular model system a Landau theory for DQPTs has been derived from microscopics~\cite{2018PhRvB..97q4303T}, for a general understanding of DQPTs it will be of central importance to explore ways that can approach DQPTs from a macroscopic perspective.
As emphasized also before, the theory of DQPTs is facing a major challenge in this context, since it is the central property of the considered nonequilibrium quantum states, that they defy a thermodynamic description and therefore a description in terms a free energy in an equilibrium sense.
On the other hand, this might not only be seen as an obstacle but rather as the defining feature of these quantum states which grant them potentially new properties, but might require a redefined notion of free energies.

\subsection{Robustness}
While DQPTs have been initially mostly studied for exactly solvable models, it has been soon recognized as an important question to which extent they are robust against perturbations~\cite{2013PhRvB..87s5104K, 2014PhRvB..90l5106K, 2015PhRvB..92j4306S}, in particular, those perturbations which break the exact solvability and make the models ergodic.
From the study of individual model systems, the phenomenology of equilibrium transitions has been recovered in the sense that DQPTs appear to be stable against weak symmetry-preserving perturbations~\cite{2013PhRvB..87s5104K, 2014PhRvB..90l5106K, 2015PhRvB..92j4306S} on the accessible transient time scales.
It is, however, not known how such perturbations can influence DQPTs occurring on long time scales, where weak perturbations can change significantly the dynamics by making for instance a nonergodic system thermalizing~\cite{2016AdPhy..65..239D}.
For the previously discussed case of a transverse-field Ising chain, where the DQPT is associated with scaling and universality, such a robustness is the consequence of an unstable RG fixed point~\cite{2015PhRvL.115n0602H}.
Upon including a symmetry-breaking perturbation, however, the character of the fixed point can change and the DQPT can then be transformed to a first-order transition without divergent correlation length~\cite{2018PhRvB..97q4303T}.
For some models it has been reported that symmetry-breaking perturbations can even lead to a smoothing of DQPTs~\cite{2010PhRvE..81b0101P}, although this might not be a general rule~\cite{2013PhRvB..87s5104K,2018PhRvB..97q4303T}.

\section{Prospect}
Summarizing, the field of DQPTs has advanced significantly in recent years.
Nevertheless, it still awaits challenges.
While some have been already pointed throughout the prior presentation, it will be the purpose in the following to discuss further open questions and prospects as well as potential future research directions.
As highlighted before, it is currently unclear to which extent DQPTs can be captured from a macroscopic perspective.
In different words: Is it possible to describe the main properties in the vicinity of a DQPT by concepts analogous to a Landau or effective field theory, which requires as an input only a few macroscopic properties such as symmetries or dimensionality?
This might also be important for classifying DQPTs on a general level, thereby extending previous approaches~\cite{2015PhRvB..91o5127V,2014PhRvL.113z5702C}.
A further intriguing prospect is to study DQPTs in 2D and 3D beyond the 1D cases addressed mostly up to now, where novel critical phenomena might appear.
This expectation is driven mainly from the equilibrium perspective, where compared to 1D it is now possible to have, for example, phase transitions at nonzero temperature or fractional and irrational critical exponents.
How these more complex critical properties express themselves in the context of DQPTs is largely unknown, which is mainly due to a methodological challenge. 
Accessing quantum dynamics in such higher dimensional systems by itself is already difficult. 
Further, on a technical level Loschmidt amplitudes or echos share the complexity of full partition functions at complex parameters, whose calculation is in addition much more demanding in most cases than determining local observables or correlation functions.
However, for example, projected-entangled pair states (PEPS) promise to provide access to 2D systems both in and out of equilibrium~\cite{2008AdPhy..57..143V}.
In this context it is important to emphasize a central advantage of DQPTs in that they occur at transient and intermediate time scales.
For PEPS for instance this implies that the entanglement production can be still limited to a tractable extent.
Another promising approach for studying quantum dynamics in 2D and 3D is to use quantum many-body state encodings on the basis of classical~\cite{2018ScPP....4...13S,2018arXiv181004178D} or artificial neural networks~\cite{2017Sci...355..602C}.
Here, however, it is not known, in general, how to compute Loschmidt amplitudes except for specific cases~\cite{2018ScPP....4...13S}.

The theory of DQPTs exhibits also important open questions in their physical interpretation.
As discussed already before in these Perspectives, DQPTs reflect a drastic change in the properties of the time-evolution operator without, however, being specific about what that change implies on physical grounds.
In equilibrium, phase transitions are characterized by order parameters marking the two phases separated by the transition.
While dynamical order parameters of topological nature have been formulated for some specific problems recently~\cite{2016PhRvB..93h5416B,2018NatPh..14..265F,2017PhRvB..96a4302B,2017PhRvB..96r0303B,2017PhRvB..96r0304H,2018arXiv180807885Z}, this might be rather seen as a first step towards a more general understanding, since order parameters are still missing for a multitude of observed DQPTs.
Thus, in the future it will be particularly important to develop new approaches for characterizing the two phases separated by a DQPT.
One promising route might be to utilize recent progress in applying machine learning approaches to quantum many-body problems, i.e., quantum phase recognition~\cite{2017PhRvX...7c1038C,2017NatSR...7.8823B}, which are not only limited to equilibrium states.

\acknowledgments
The author is thankful for the stimulating discussions and collaborations on topics related to this review with various colleagues including Debasish Banerjee, Rainer Blatt, Jan Budich, Sebastian Diehl, Nick Fläschner, Philipp Hauke, Yi-Ping Huang, Petar Jurcevic, Ben Lanyon, Achilleas Lazarides, Stefan Kehrein, Michael Knap, Marcus Kollar, Roderich Moessner, Anatoli Polkovnikov, Christian Roos, Markus Schmitt, Alessandro Silva, Daniele Trapin, Dominik Vogel, Matthias
Vojta, Simon Weidinger, Christof Weitenberg, Peter Zoller, and Bojan Zunkovic. Financial support by the Deutsche Forschungsgemeinschaft via the Gottfried Wilhelm Leibniz Prize program is gratefully acknowledged.

\bibliographystyle{eplbib}
\bibliography{literature.bib}

\begin{thebibliography}{10}
\expandafter\ifx\csname url\endcsname\relax\def\url#1{\texttt{#1}}\fi

\bibitem{1987imsm.book.....C}
\Name{{Chandler} D.} \Book{{Introduction to Modern Statistical Mechanics}}
  1987.

\bibitem{1998AmJPh..66..164C}
\Name{{Callen} H.~B. \and {Scott} H.~L.} \REVIEW{American Journal of
  Physics}{66}{1998}{164}.

\bibitem{2011qpt..book.....S}
\Name{{Sachdev} S.} \Book{{Quantum Phase Transitions}} 2011.

\bibitem{2013PhRvL.110m5704H}
\Name{{Heyl} M., {Polkovnikov} A. \and {Kehrein} S.}
  \REVIEW{\prl}{110}{2013}{135704}.

\bibitem{2018RPPh...81e4001H}
\Name{{Heyl} M.} \REVIEW{Reports on Progress in Physics}{81}{2018}{054001}.

\bibitem{2017PhRvL.119h0501J}
\Name{{Jurcevic} P., {Shen} H., {Hauke} P., {Maier} C., {Brydges} T., {Hempel}
  C., {Lanyon} B.~P., {Heyl} M., {Blatt} R. \and {Roos} C.~F.}
  \REVIEW{\prl}{119}{2017}{080501}.

\bibitem{2011RvMP...83..863P}
\Name{{Polkovnikov} A., {Sengupta} K., {Silva} A. \and {Vengalattore} M.}
  \REVIEW{Reviews of Modern Physics}{83}{2011}{863}.

\bibitem{2015NatPh..11..124E}
\Name{{Eisert} J., {Friesdorf} M. \and {Gogolin} C.} \REVIEW{Nature
  Physics}{11}{2015}{124}.

\bibitem{2015ARCMP...6..383A}
\Name{{Altman} E. \and {Vosk} R.} \REVIEW{Annual Review of Condensed Matter
  Physics}{6}{2015}{383}.

\bibitem{2015ARCMP...6...15N}
\Name{{Nandkishore} R. \and {Huse} D.~A.} \REVIEW{Annual Review of Condensed
  Matter Physics}{6}{2015}{15}.

\bibitem{2017NatPh..13..424M}
\Name{{Moessner} R. \and {Sondhi} S.~L.} \REVIEW{Nature
  Physics}{13}{2017}{424}.

\bibitem{2008RvMP...80..885B}
\Name{{Bloch} I., {Dalibard} J. \and {Zwerger} W.} \REVIEW{Reviews of Modern
  Physics}{80}{2008}{885}.

\bibitem{2012NatPh...8..267B}
\Name{{Bloch} I., {Dalibard} J. \and {Nascimb{\`e}ne} S.} \REVIEW{Nature
  Physics}{8}{2012}{267}.

\bibitem{2012NatPh...8..277B}
\Name{{Blatt} R. \and {Roos} C.~F.} \REVIEW{Nature Physics}{8}{2012}{277}.

\bibitem{2014RvMP...86..153G}
\Name{{Georgescu} I.~M., {Ashhab} S. \and {Nori} F.} \REVIEW{Reviews of Modern
  Physics}{86}{2014}{153}.

\bibitem{2015Sci...349..842S}
\Name{{Schreiber} M., {Hodgman} S.~S., {Bordia} P., {L{\"u}schen} H.~P.,
  {Fischer} M.~H., {Vosk} R., {Altman} E., {Schneider} U. \and {Bloch} I.}
  \REVIEW{Science}{349}{2015}{842}.

\bibitem{2016Sci...352.1547C}
\Name{{Choi} J.-y., {Hild} S., {Zeiher} J., {Schau{\ss}} P., {Rubio-Abadal} A.,
  {Yefsah} T., {Khemani} V., {Huse} D.~A., {Bloch} I. \and {Gross} C.}
  \REVIEW{Science}{352}{2016}{1547}.

\bibitem{2016NatPh..12..907S}
\Name{{Smith} J., {Lee} A., {Richerme} P., {Neyenhuis} B., {Hess} P.~W.,
  {Hauke} P., {Heyl} M., {Huse} D.~A. \and {Monroe} C.} \REVIEW{Nature
  Physics}{12}{2016}{907}.

\bibitem{2012Sci...337.1318G}
\Name{{Gring} M., {Kuhnert} M., {Langen} T., {Kitagawa} T., {Rauer} B.,
  {Schreitl} M., {Mazets} I., {Smith} D.~A., {Demler} E. \and {Schmiedmayer}
  J.} \REVIEW{Science}{337}{2012}{1318}.

\bibitem{2017SciA....3E0672N}
\Name{{Neyenhuis} B., {Zhang} J., {Hess} P.~W., {Smith} J., {Lee} A.~C.,
  {Richerme} P., {Gong} Z.-X., {Gorshkov} A.~V. \and {Monroe} C.}
  \REVIEW{Science Advances}{3}{2017}{e1700672}.

\bibitem{2018arXiv180905554S}
\Name{{Singh} K., {Fujiwara} K.~M., {Geiger} Z.~A., {Simmons} E.~Q., {Lipatov}
  M., {Cao} A., {Dotti} P., {Rajagopal} S.~V., {Senaratne} R., {Shimasaki} T.,
  {Heyl} M., {Eckardt} A. \and {Weld} D.~M.} \REVIEW{ArXiv
  e-prints}{}{2018}{arXiv:1809.05554}.

\bibitem{2016Natur.534..516M}
\Name{{Martinez} E.~A., {Muschik} C.~A., {Schindler} P., {Nigg} D., {Erhard}
  A., {Heyl} M., {Hauke} P., {Dalmonte} M., {Monz} T., {Zoller} P. \and {Blatt}
  R.} \REVIEW{\nat}{534}{2016}{516}.

\bibitem{2017Natur.543..221C}
\Name{{Choi} S., {Choi} J., {Landig} R., {Kucsko} G., {Zhou} H., {Isoya} J.,
  {Jelezko} F., {Onoda} S., {Sumiya} H., {Khemani} V., {von Keyserlingk} C.,
  {Yao} N.~Y., {Demler} E. \and {Lukin} M.~D.} \REVIEW{\nat}{543}{2017}{221}.

\bibitem{2017Natur.543..217Z}
\Name{{Zhang} J., {Hess} P.~W., {Kyprianidis} A., {Becker} P., {Lee} A.,
  {Smith} J., {Pagano} G., {Potirniche} I.~D., {Potter} A.~C., {Vishwanath} A.,
  {Yao} N.~Y. \and {Monroe} C.} \REVIEW{\nat}{543}{2017}{217}.

\bibitem{2018NatPh..14..265F}
\Name{{Fl{\"a}schner} N., {Vogel} D., {Tarnowski} M., {Rem} B.~S.,
  {L{\"u}hmann} D.~S., {Heyl} M., {Budich} J.~C., {Mathey} L., {Sengstock} K.
  \and {Weitenberg} C.} \REVIEW{Nature Physics}{14}{2018}{265}.

\bibitem{2017Natur.551..601Z}
\Name{{Zhang} J., {Pagano} G., {Hess} P.~W., {Kyprianidis} A., {Becker} P.,
  {Kaplan} H., {Gorshkov} A.~V., {Gong} Z.~X. \and {Monroe} C.}
  \REVIEW{\nat}{551}{2017}{601}.

\bibitem{2018arXiv180609269G}
\Name{{Guo} X.-Y., {Yang} C., {Zeng} Y., {Peng} Y., {Li} H.-K., {Deng} H.,
  {Jin} Y.-R., {Chen} S., {Zheng} D. \and {Fan} H.} \REVIEW{ArXiv
  e-prints}{}{2018}{arXiv:1806.09269}.

\bibitem{2018arXiv180704483T}
\Name{{Tian} T., {Ke} Y., {Zhang} L., {Lin} S., {Shi} Z., {Huang} P., {Lee} C.
  \and {Du} J.} \REVIEW{ArXiv e-prints}{}{2018}{arXiv:1807.04483}.

\bibitem{2018arXiv180610871W}
\Name{{Wang} K., {Qiu} X., {Xiao} L., {Zhan} X., {Bian} Z., {Yi} W. \and {Xue}
  P.} \REVIEW{ArXiv e-prints}{}{2018}{arXiv:1806.10871}.

\bibitem{2018arXiv180803930X}
\Name{{Xu} X.-Y., {Wang} Q.-Q., {Heyl} M., {Budich} J.~C., {Pan} W.-W., {Chen}
  Z., {Jan} M., {Sun} K., {Xu} J.-S., {Han} Y.-J., {Li} C.-F. \and {Guo} G.-C.}
  \REVIEW{ArXiv e-prints}{}{2018}{arXiv:1808.03930}.

\bibitem{2016PhRvL.116y0401K}
\Name{{Khemani} V., {Lazarides} A., {Moessner} R. \and {Sondhi} S.~L.}
  \REVIEW{\prl}{116}{2016}{250401}.

\bibitem{2016PhRvL.117i0402E}
\Name{{Else} D.~V., {Bauer} B. \and {Nayak} C.}
  \REVIEW{\prl}{117}{2016}{090402}.

\bibitem{2013PhRvL.110k8901B}
\Name{{Bruno} P.} \REVIEW{\prl}{110}{2013}{118901}.

\bibitem{2015PhRvL.114y1603W}
\Name{{Watanabe} H. \and {Oshikawa} M.} \REVIEW{\prl}{114}{2015}{251603}.

\bibitem{2016PhRvB..93j4302A}
\Name{{Abeling} N.~O. \and {Kehrein} S.} \REVIEW{Physical Review
  B}{93}{2016}{104302}.

\bibitem{2017PhRvB..96r0303B}
\Name{{Bhattacharya} U., {Bandyopadhyay} S. \and {Dutta} A.} \REVIEW{Physical
  Review B}{96}{2017}{180303}.

\bibitem{2017PhRvB..96r0304H}
\Name{{Heyl} M. \and {Budich} J.~C.} \REVIEW{Physical Review
  B}{96}{2017}{180304}.

\bibitem{2018PhRvB..97q4401L}
\Name{{Lang} J., {Frank} B. \and {Halimeh} J.~C.} \REVIEW{Physical Review
  B}{97}{2018}{174401}.

\bibitem{2018NatSR...811921B}
\Name{{Bandyopadhyay} S., {Laha} S., {Bhattacharya} U. \and {Dutta} A.}
  \REVIEW{Scientific Reports}{8}{2018}{11921}.

\bibitem{2010PhRvE..81b0101P}
\Name{{Pollmann} F., {Mukerjee} S., {Green} A.~G. \and {Moore} J.~E.}
  \REVIEW{\pre}{81}{2010}{020101}.

\bibitem{1995NuPhB.453..581L}
\Name{{LeClair} A., {Mussardo} G., {Saleur} H. \and {Skorik} S.}
  \REVIEW{Nuclear Physics B}{453}{1995}{581}.

\bibitem{2003PhRvL..90v7902V}
\Name{{Vidal} G., {Latorre} J.~I., {Rico} E. \and {Kitaev} A.}
  \REVIEW{\prl}{90}{2003}{227902}.

\bibitem{2014PhRvB..89p1105V}
\Name{{Vajna} S. \and {D{\'o}ra} B.} \REVIEW{Physical Review
  B}{89}{2014}{161105}.

\bibitem{2014PhRvB..89l5120A}
\Name{{Andraschko} F. \and {Sirker} J.} \REVIEW{Physical Review
  B}{89}{2014}{125120}.

\bibitem{2015PhRvB..92g5114S}
\Name{{Schmitt} M. \and {Kehrein} S.} \REVIEW{Physical Review
  B}{92}{2015}{075114}.

\bibitem{2017PhRvB..96m4427H}
\Name{{Halimeh} J.~C. \and {Zauner-Stauber} V.} \REVIEW{Physical Review
  B}{96}{2017}{134427}.

\bibitem{2015PhRvB..91o5127V}
\Name{{Vajna} S. \and {D{\'o}ra} B.} \REVIEW{Physical Review
  B}{91}{2015}{155127}.

\bibitem{2016PhRvB..93h5416B}
\Name{{Budich} J.~C. \and {Heyl} M.} \REVIEW{Physical Review
  B}{93}{2016}{085416}.

\bibitem{2016PhRvL.117h6802H}
\Name{{Huang} Z. \and {Balatsky} A.~V.} \REVIEW{\prl}{117}{2016}{086802}.

\bibitem{2014PhRvL.113z5702C}
\Name{{Canovi} E., {Werner} P. \and {Eckstein} M.}
  \REVIEW{\prl}{113}{2014}{265702}.

\bibitem{2015PhRvL.115n0602H}
\Name{{Heyl} M.} \REVIEW{\prl}{115}{2015}{140602}.

\bibitem{2018PhRvL.120m0601Z}
\Name{{{\v{Z}}unkovi{\v{c}}} B., {Heyl} M., {Knap} M. \and {Silva} A.}
  \REVIEW{\prl}{120}{2018}{130601}.

\bibitem{2017PhRvB..96m4313W}
\Name{{Weidinger} S.~A., {Heyl} M., {Silva} A. \and {Knap} M.} \REVIEW{Physical
  Review B}{96}{2017}{134313}.

\bibitem{2018arXiv180807874H}
\Name{{Huang} Y.-P., {Banerjee} D. \and {Heyl} M.} \REVIEW{ArXiv
  e-prints}{}{2018}{arXiv:1808.07874}.

\bibitem{2014PhRvL.113t5701H}
\Name{{Heyl} M.} \REVIEW{\prl}{113}{2014}{205701}.

\bibitem{2017PhRvB..96a4302B}
\Name{{Bhattacharya} U. \and {Dutta} A.} \REVIEW{Physical Review
  B}{96}{2017}{014302}.

\bibitem{2018arXiv180807885Z}
\Name{{Zache} T.~V., {Mueller} N., {Schneider} J.~T., {Jendrzejewski} F.,
  {Berges} J. \and {Hauke} P.} \REVIEW{ArXiv
  e-prints}{}{2018}{arXiv:1808.07885}.

\bibitem{1972CMaPh..28..251L}
\Name{{Lieb} E.~H. \and {Robinson} D.~W.} \REVIEW{Communications in
  Mathematical Physics}{28}{1972}{251}.

\bibitem{2006JSP...124....1N}
\Name{{Nachtergaele} B., {Ogata} Y. \and {Sims} R.} \REVIEW{Journal of
  Statistical Physics}{124}{2006}{1}.

\bibitem{2006CMaPh.265..781H}
\Name{{Hastings} M.~B. \and {Koma} T.} \REVIEW{Communications in Mathematical
  Physics}{265}{2006}{781}.

\bibitem{1992cond.mat..4009S}
\Name{{Sethna} J.~P.} \REVIEW{ArXiv e-prints}{}{1992}{cond}.

\bibitem{2018ScPP....4...13S}
\Name{{Schmitt} M. \and {Heyl} M.} \REVIEW{SciPost Physics}{4}{2018}{013}.

\bibitem{2018PhRvB..97q4303T}
\Name{{Trapin} D. \and {Heyl} M.} \REVIEW{Physical Review B}{97}{2018}{174303}.

\bibitem{2013PhRvB..87s5104K}
\Name{{Karrasch} C. \and {Schuricht} D.} \REVIEW{Physical Review
  B}{87}{2013}{195104}.

\bibitem{2014PhRvB..90l5106K}
\Name{{Kriel} J.~N., {Karrasch} C. \and {Kehrein} S.} \REVIEW{Physical Review
  B}{90}{2014}{125106}.

\bibitem{2015PhRvB..92j4306S}
\Name{{Sharma} S., {Suzuki} S. \and {Dutta} A.} \REVIEW{Physical Review
  B}{92}{2015}{104306}.

\bibitem{2016AdPhy..65..239D}
\Name{{D'Alessio} L., {Kafri} Y., {Polkovnikov} A. \and {Rigol} M.}
  \REVIEW{Advances in Physics}{65}{2016}{239}.

\bibitem{2008AdPhy..57..143V}
\Name{{Verstraete} F., {Murg} V. \and {Cirac} J.~I.} \REVIEW{Advances in
  Physics}{57}{2008}{143}.

\bibitem{2018arXiv181004178D}
\Name{{De Tomasi} G., {Pollmann} F. \and {Heyl} M.} \REVIEW{ArXiv
  e-prints}{}{2018}{arXiv:1810.04178}.

\bibitem{2017Sci...355..602C}
\Name{{Carleo} G. \and {Troyer} M.} \REVIEW{Science}{355}{2017}{602}.

\bibitem{2017PhRvX...7c1038C}
\Name{{Ch'ng} K., {Carrasquilla} J., {Melko} R.~G. \and {Khatami} E.}
  \REVIEW{Physical Review X}{7}{2017}{031038}.

\bibitem{2017NatSR...7.8823B}
\Name{{Broecker} P., {Carrasquilla} J., {Melko} R.~G. \and {Trebst} S.}
  \REVIEW{Scientific Reports}{7}{2017}{8823}.

\end{thebibliography}

\end{document}